\documentclass[twocolumn,apjl]{aastex61}
\usepackage{ifthen}
\usepackage{etoolbox}

\usepackage{import}
\usepackage{longtable}
\usepackage{graphicx}
\usepackage{amsmath,amssymb}
\usepackage{color}
\usepackage{units}
\usepackage{epstopdf}
\usepackage{hyperref}
\usepackage{multirow}
\usepackage[caption=false]{subfig}
\usepackage{float}

\usepackage{import}
\usepackage{longtable}
\usepackage{graphicx}
\usepackage{amsmath,amssymb}
\usepackage{color}
\usepackage{units}
\usepackage{epstopdf}
\usepackage{hyperref}
\usepackage{multirow}
\usepackage[caption=false]{subfig}
\usepackage{float}

\newcommand{\beq}{\begin{equation}}
\newcommand{\eeq}{\end{equation}}
\newcommand{\bdm}{\begin{displaymath}}
\newcommand{\edm}{\end{displaymath}}

\definecolor{Gray}{gray}{0.9}
\definecolor{orange}{rgb}{0.9,0.5,0}

\newcommand*\diff{\mathop{}\!\mathrm{d}}

\newcommand{\dist}{{\mathcal D}}

\newcommand{\icecube}[0]{IceCube}

\newcommand{\Msun}{\>{\rm M_{\odot}}}

\def\newacronym#1#2#3{\gdef#1{#3 (#2)\gdef#1{#2}}}

\newacronym{\gt}{Georgia Tech}{Georgia Institute of Technology}
\newacronym{\icecube}{IceCube}{IceCube South Pole Neutrino Observatory}
\newacronym{\cta}{CTA}{Cherenkov Telescope Array}
\newacronym{\sdsc}{SDSC}{San Diego Supercomputer Center}
\newacronym{\cra}{CRA}{Center for Relativistic Astrophysics}
\newacronym{\nr}{NR}{numerical relativity}
\newacronym{\ornl}{ORNL}{Oak Ridge National Laboratory}
\newacronym{\lisa}{LISA}{Laser Interferometer Space Antenna}
\newacronym{\hst}{HST}{Hubble Space Telescope}
\newacronym{\jwst}{JWST}{James Webb Space Telescope}
\newacronym{\alma}{ALMA}{Atacama Large Millimeter Array}
\newacronym{\ligo}{LIGO}{Laser Interferometer Gravitational Wave Observatory}
\newacronym{\sph}{SPH}{smooth particle hydrodynamics}
\newacronym{\tsi}{TSI}{Terascale Supernova Initiative}
\newacronym{\wmap}{WMAP}{the Wilkinson Microwave Anisotropy Probe}
\newacronym{\cmb}{CMB}{cosmic microwave background}
\newacronym{\ibbh}{IBBH}{intermediate binary black hole}
\newacronym{\grb}{GRB}{Gamma-ray burst}
\newacronym{\grbs}{GRBs}{Gamma-ray bursts}
\newacronym{\hpc}{HPC}{High Performance Computing}
\newacronym{\htc}{HTC}{High Throughput Computing}
\newacronym{\bssn}{BSSN}{Baumgarte-Shapiro-Shibata-Nakamura}
\newacronym{\amr}{AMR}{adaptive mesh refinement}
\newacronym{\mpi}{MPI}{Message Passing Interface}
\newacronym{\cuda}{CUDA}{Compute Unified Device Architecture}
\newacronym{\gui}{GUI}{Graphical User Inferface}
\newacronym{\oit}{OIT}{Office of Information Technologies}
\newacronym{\lbg}{LBGs}{Lyman break galaxies}
\newacronym{\mhd}{MHD}{magneto-hydrodynamics}
\newacronym{\imf}{IMF}{initial mass function}
\newacronym{\sms}{SMS}{supermassive star}
\newacronym{\posix}{POSIX}{Portable Operating System Interface}
\newacronym{\ssd}{SSD}{Solid State Drive}
\newacronym{\os}{OS}{Operating System}
\newacronym{\pace}{PACE}{Partnership for an Advanced Computing Environment}
\newacronym{\oit}{OIT}{Office of Information Technology}
\newacronym{\gtswd}{GTSWD}{Georgia Tech Software Distribution}
\newacronym{\uga}{UGA}{University of Georgia}
\newacronym{\mahpic}{MaHPIC}{Malaria Host-Pathogen Interaction Center}
\newacronym{\nih}{NIH}{National Institutes of Health}
\newacronym{\cdc}{CDC}{Centers for Disease Control and Prevention}
\newacronym{\xsede}{XSEDE}{Extreme Science and Engineering Discovery Environment}
\newacronym{\osg}{OSG}{Open Science Grid}
\newacronym{\egi}{EGI}{European Grid Infrastructure}
\newacronym{\vo} {VO}{Virtual Organization}
\newacronym{\sox}{SoX}{Southern Crossroads}
\newacronym{\slr}{SLR}{Southern Light Rail}
\newacronym{\ddn}{DDN}{DataDirect Networks}
\newacronym{\rsv}{RSV}{Resource and Service Validation}
\newacronym{\cvmfs}{CVMFS}{CernVM File System}
\newacronym{\nfs}{NFS}{Network File Sytem}
\newacronym{\lsc}{LSC}{LIGO Scientific Collaboration}

\newacronym{\DA}{DA}{data analysis}
\newacronym{\CBC}{CBC}{Compact Binary Coalescences}
\newacronym{\lvc}{LVC}{LIGO-Virgo Collaboration}
\newacronym{\aligo}{aLIGO}{Advanced LIGO}
\newacronym{\ldr}{LDR}{LIGO Data Replicator}

\def\si#1{Senior Investigator#1 (SI#1)\gdef\si{SI}}
\def\emri#1{Extreme Mass-Ratio Inspiral#1 (EMRI#1)\gdef\emri{EMRI}}
\def\imbh#1{intermediate mass black hole#1 (IMBH#1)\gdef\imbh{IMBH}}
\def\smbh#1{supermassive black hole#1 (SMBH#1)\gdef\smbh{SMBH}}

\def\bbh#1{binary black hole#1 (BBH#1)\gdef\bbh{BBH}}
\def\bh#1{black hole#1 (BH#1)\gdef\bh{BH}}
\def\ns#1{neutron star#1 (NS#1)\gdef\ns{NS}}
\def\bns#1{binary neutron star#1 (BNS#1)\gdef\bns{BNS}}
\def\nsbh#1{neutron star--black hole#1 (NSBH#1)\gdef\nsbh{NSBH}}
\def\hmns#1{hypermassive neutron star#1 (HMNS#1)\gdef\hmns{HMNS}}
\def\gw#1{gravitational wave#1 (GW#1)\gdef\gw{GW}}
\def\grb#1{Gamma-ray burst#1 (GRB#1)\gdef\grb{GRB}}
\def\pnw#1{post-Newtonian#1 (PN#1)\gdef\pnw{PN}}
\def\eos#1{equation of state#1 (EOS#1)\gdef\eos{EOS}}
\def\gpu#1{graphics processing unit#1 (GPU#1)\gdef\gpu{GPU}}
\def\sn#1{supernova#1 (SN#1)\gdef\sn{SN}}
\def\lae#1{Lyman Alpha Emitter#1 (LAE#1)\gdef\lae{LAE}}
\def\sfr#1{star formation rate#1 (SFR#1)\gdef\sfr{SFR}}
\def\sed#1{spectral energy density#1 (SED#1)\gdef\sed{SED}}
\def\lsst#1{Large Synaptic Survey Telescope#1 (LSST#1)\gdef\lsst{LSST}}
\def\rproc#1{rapid neutron capture#1 ($r$-process#1)\gdef\rproc{$r$-process#1}}

\newcommand{\code}[1]{\textsl{#1}}

\newacronym{\grhd}{GRHD}{general relativistic hydrodynamics}
\newacronym{\grmhd}{GRMHD}{general relativistic magnetohydrodynamics}
\newacronym{\gw}{GW}{gravitational wave}
\newacronym{\EM}{EM}{electromagnetic}
\newacronym{\utb}{UTB}{University of Texas (Brownsville)}
\newacronym{\aps}{APS}{American Physical Society}
\newacronym{\lsu}{LSU}{Louisiana State University}
\newacronym{\rit}{RIT}{Rochester Institute of Technology}
\newcommand{\nn}{\nonumber}


\newcommand{\macro}[1]{#1}

\newcommand{\magnetarULSTAMP}{\macro{\ensuremath{4\,\Msun c^2}}}
\newcommand{\barmodesULSTAMP}{\macro{\ensuremath{2\,\Msun c^2}}}
\newcommand{\barmodesALonetrack}{\macro{\ensuremath{8.4}}}
\newcommand{\barmodesBLonetrack}{\macro{\ensuremath{9.0}}}
\newcommand{\barmodesCLonetrack}{\macro{\ensuremath{6.5}}}
\newcommand{\barmodesDLonetrack}{\macro{\ensuremath{8.1}}}
\newcommand{\barmodesELonetrack}{\macro{\ensuremath{8.3}}}
\newcommand{\barmodesFLonetrack}{\macro{\ensuremath{5.9}}}
\newcommand{\msmagnetarALonetrack}{\macro{\ensuremath{9.7}}}
\newcommand{\msmagnetarBLonetrack}{\macro{\ensuremath{8.4}}}
\newcommand{\msmagnetarCLonetrack}{\macro{\ensuremath{10}}}
\newcommand{\msmagnetarDLonetrack}{\macro{\ensuremath{9.8}}}
\newcommand{\msmagnetarILonetrack}{\macro{\ensuremath{15}}}
\newcommand{\msmagnetarJLonetrack}{\macro{\ensuremath{16}}}
\newcommand{\msmagnetarKLonetrack}{\macro{\ensuremath{26}}}
\newcommand{\msmagnetarLLonetrack}{\macro{\ensuremath{21}}}
\newcommand{\barmodesAZebragard}{\macro{\ensuremath{10}}}
\newcommand{\barmodesBZebragard}{\macro{\ensuremath{9.2}}}
\newcommand{\barmodesCZebragard}{\macro{\ensuremath{7.6}}}
\newcommand{\barmodesDZebragard}{\macro{\ensuremath{10}}}
\newcommand{\barmodesEZebragard}{\macro{\ensuremath{11}}}
\newcommand{\barmodesFZebragard}{\macro{\ensuremath{7.9}}}
\newcommand{\msmagnetarAZebragard}{\macro{\ensuremath{11}}}
\newcommand{\msmagnetarBZebragard}{\macro{\ensuremath{11}}}
\newcommand{\msmagnetarCZebragard}{\macro{\ensuremath{23}}}
\newcommand{\msmagnetarDZebragard}{\macro{\ensuremath{24}}}
\newcommand{\msmagnetarIZebragard}{\macro{\ensuremath{28}}}
\newcommand{\msmagnetarJZebragard}{\macro{\ensuremath{27}}}
\newcommand{\msmagnetarKZebragard}{\macro{\ensuremath{35}}}
\newcommand{\msmagnetarLZebragard}{\macro{\ensuremath{35}}}
\newcommand{\barmodesAcWB}{\macro{\ensuremath{8.3}}}
\newcommand{\barmodesBcWB}{\macro{\ensuremath{8.3}}}
\newcommand{\barmodesCcWB}{\macro{\ensuremath{7.6}}}
\newcommand{\barmodesDcWB}{\macro{\ensuremath{7.9}}}
\newcommand{\barmodesEcWB}{\macro{\ensuremath{8.1}}}
\newcommand{\barmodesFcWB}{\macro{\ensuremath{6.2}}}

\newcommand{\progMassLow}{\macro{\ensuremath{2.73\,}}}
\newcommand{\progMassHigh}{\macro{\ensuremath{3.29\,}}}
\newcommand{\distance}{\macro{\ensuremath{40_{-14}^{+8}\,}}}
\newcommand{\distanceHubble}{\macro{\ensuremath{43.8_{-6.9}^{+2.9}\,}}}
\newcommand{\FermiDelay}{\macro{\ensuremath{1.7\,}}}

\newtoggle{endauthorlist}
\toggletrue{endauthorlist}

\begin{document}

\title{Search for post-merger gravitational waves from the remnant of the binary neutron star merger GW170817}


\iftoggle{endauthorlist}{
  %
  %
  \let\mymaketitle\maketitle
  \let\myauthor\author
  \let\myaffiliation\affiliation
  \author{The LIGO Scientific Collaboration}
  \author{The Virgo Collaboration}
}{
  %
  %
  \author{B.~P.~Abbott}
\affiliation{LIGO, California Institute of Technology, Pasadena, CA 91125, USA}
\author{R.~Abbott}
\affiliation{LIGO, California Institute of Technology, Pasadena, CA 91125, USA}
\author{T.~D.~Abbott}
\affiliation{Louisiana State University, Baton Rouge, LA 70803, USA}
\author{F.~Acernese}
\affiliation{Universit\`a di Salerno, Fisciano, I-84084 Salerno, Italy}
\affiliation{INFN, Sezione di Napoli, Complesso Universitario di Monte S.Angelo, I-80126 Napoli, Italy}
\author{K.~Ackley}
\affiliation{University of Florida, Gainesville, FL 32611, USA}
\affiliation{OzGrav, School of Physics \& Astronomy, Monash University, Clayton 3800, Victoria, Australia}
\author{C.~Adams}
\affiliation{LIGO Livingston Observatory, Livingston, LA 70754, USA}
\author{T.~Adams}
\affiliation{Laboratoire d'Annecy-le-Vieux de Physique des Particules (LAPP), Universit\'e Savoie Mont Blanc, CNRS/IN2P3, F-74941 Annecy, France}
\author{P.~Addesso}
\affiliation{University of Sannio at Benevento, I-82100 Benevento, Italy and INFN, Sezione di Napoli, I-80100 Napoli, Italy}
\author{R.~X.~Adhikari}
\affiliation{LIGO, California Institute of Technology, Pasadena, CA 91125, USA}
\author{V.~B.~Adya}
\affiliation{Max Planck Institute for Gravitational Physics (Albert Einstein Institute), D-30167 Hannover, Germany}
% ...
\collaboration{LIGO Scientific Collaboration and Virgo Collaboration}
}

\begin{abstract}
The first observation of a binary neutron star coalescence by the Advanced LIGO and Advanced Virgo gravitational-wave detectors offers an unprecedented opportunity to study matter under the most extreme conditions.
After such a merger, a compact remnant is left over whose nature depends primarily on the masses of the inspiralling objects and on the equation of state of nuclear matter.
This could be either a black hole or a neutron star (NS),
with the latter being either long-lived or too massive for stability implying delayed collapse to a black hole.
Here, we present a search for gravitational waves from the remnant of the binary neutron star merger GW170817 using data from Advanced LIGO and Advanced Virgo. 
We search for short ($\lesssim1$\,s) and intermediate-duration ($\lesssim500$\,s) signals,
which includes gravitational-wave emission from a hypermassive NS or supramassive NS, respectively.
We find no signal from the  post-merger remnant.  Our derived strain upper limits are 
more than an order of magnitude larger than those predicted by most models.
For short signals, our best upper limit on the
root-sum-square of the gravitational-wave strain emitted from 1--4\,kHz is $h_{\rm rss}^{50\%}=2.1\times 10^{-22}$\,Hz$^{-1/2}$
at 50\% detection efficiency.
For intermediate-duration signals, our best upper limit at 50\% detection efficiency
is $h_{\rm rss}^{50\%}=\msmagnetarBLonetrack\times 10^{-22}$\,Hz$^{-1/2}$ for a millisecond magnetar model, 
and $h_{\rm rss}^{50\%}=\barmodesFLonetrack\times 10^{-22}$\,Hz$^{-1/2}$ for a bar-mode model.
These results indicate that post-merger emission from a similar event may be detectable when advanced detectors reach design sensitivity or with next-generation detectors.
\end{abstract}


\section{Introduction}
\label{sec:Intro}
On August 17, 2017 12:41:04.4 UTC, the two detectors of the Advanced Laser Interferometer Gravitational-Wave Observatory (LIGO) and the Advanced Virgo detector observed GW170817, the \gw{} signal from the coalescence of two compact objects, almost certainly neutron stars (NSs) \citep{DetectionPaper}.
Supporting this hypothesis were electromagnetic counterparts observed across the spectrum~\citep{MMAPaper,GRBGWPaper}.
Thanks to its relatively close proximity to Earth, with 90\% credible intervals of \distance Mpc as measured by the GW data analysis~\citep{DetectionPaper} and \distanceHubble Mpc as measured with electromagnetic observations~\citep{AbEA2017b}, GW170817 offers the first opportunity to study the nature of the remnant leftover from a binary NS merger using GW observations. 

The merger of two NSs can have four possible outcomes:
(i) The prompt formation of a black hole (BH),
(ii) the formation of a hypermassive NS that collapses to a BH in $\lesssim1$ s,
(iii) the formation of a supramassive NS that collapses to a BH on timescales of $\sim10-10^4$ s,
or (iv) the formation of a stable NS.
The specific outcome of any merger depends on the progenitor masses, with the two NSs that merged in GW170817 having a total mass between \progMassLow\ and \progMassHigh\,$M_\odot$ (using the high-spin priors)~\citep{DetectionPaper}, and also on the NS equation of state.  We present a broad search for both short ($\lesssim 1$\,s) and intermediate ($\lesssim500$\,s) duration \gw{} signals potentially emitted from post-merger remnants in scenarios (ii), (iii) and (iv).  We find no evidence for a statistically significant signal and set upper limits on possible \gw{} strain amplitudes and \gw{} energy emission.

Before describing the search, we briefly review the four scenarios listed above.
If the system promptly forms a BH, the GW quasinormal-mode ringdown signal from a remnant BH in the GW170817 mass range has a dominant frequency around 6 kHz~\citep{ShTa2006,BaGi2008}.
Current \gw{} detectors are not robustly calibrated at such high frequencies.
Moreover, for such a remnant BH the ringdown signal-to-noise ratio at $\sim40$\,Mpc is vanishingly small.
We therefore focus on short and intermediate-duration GW signals from a possible NS remnant.
We also do not target GW emission from a delayed NS-to-BH collapse in scenarios (ii) or (iii) as
it is also not likely detectable~\citep[e.g.,][]{BaHa2007}.

A hypermassive NS is one that has mass greater than the maximum mass of a uniformly rotating star, but is prevented from collapse through support from differential rotation and thermal gradients~\citep{BaSh2000}.  Rapid cooling through neutrino emission and magnetic braking of the differential rotation causes such merger remnants to collapse $\lesssim1$ s after formation~\citep{Shap2000,HoKi2013}. 
If the star is less massive but still supramassive---i.e., its mass is larger than the maximum for a non-rotating NS---it will spin down through electromagnetic and GW emission, eventually collapsing to a BH between $\sim10$ and $5\times10^4$\,s after merger \citep{RaLa2014}. 

Taking the posterior distribution for the progenitor masses of GW170817~\citep{DetectionPaper}, one can calculate a probability distribution for the gravitational mass of the post-merger remnant assuming conservation of baryonic mass (and neglecting mass loss to the ejecta).  For a broad range of equations of state, this post-merger mass lies in the hypermassive NS regime~\citep[see Sec. 5.2 of ][]{GRBGWPaper}.

Moreover, observations of a kilonova-like counterpart in the optical and infrared can give insight into the remnant. For example, observations suggest low-lanthanide ejecta from the merger~\citep{Smartt2017}, which may be the result of a hypermassive NS surviving $\gtrsim100$ ms after the merger causing additional neutrino flux over that of prompt BH formation to irradiate the ejecta, increasing the electron fraction and not allowing the formation of lanthanides~\citep[][]{MeFa2014, GRBGWPaper, KilonovaePaper}.
However, optical observations at late times also support opacity-heavy models, potentially implying a hypermassive NS lifetime $< 100$\,ms~\citep{Smartt2017}.

A hypermassive NS remnant may also partially explain the delay between the coalescence time of GW170817 and the trigger time of the short $\gamma$-ray burst (GRB) 170817A, detected \FermiDelay \,s later by the \textit{Fermi} Gamma-ray Burst Monitor \citep{GBM_only_paper,GRBGWPaper}.

Simulations of merging binary NSs with hypermassive remnants show that the post-merger GW emission is dominated by the quadrupolar $f$-mode~\citep[$\sim2$---$4$ kHz;][]{Xing1994,1996A&A...311..532R,PhysRevD.61.064001}, with broad secondary and tertiary peaks in the $\sim 1.8$---$4$ kHz range~\citep{HoKi2013}.  Depending on the \eos{}, the \gw{} signal may include contributions from post-merger emission beginning around 1\,kHz~\citep{2017PhRvD..96f3011M}.  The structure and locations of the spectral peaks is correlated with the masses and spins of the progenitors~\citep{BeNa2014, KaGa2014, BaSt2015} and the nuclear equation of state \citep{ReBa2013,BeDi2015,ReTa2016}, implying GW observations of a hypermassive NS potentially enable strong constraints on the equation of state~\citep{Shib2005,Bauswein2011}.  
 
We also consider the scenarios (iii) and (iv) of a longer-lived post-merger remnant.
Observations of X-ray afterglows following short GRBs indicate that a fraction of binary NS mergers may result in supramassive or stable NSs lasting $\gg100$\,s \cite[e.g.,][]{Row2013,Lu2015}.
GRB\,170817A was sub-energetic compared to the population of cosmological short GRBs \citep{Be2014,MMAPaper,GBM_only_paper}, had an atypical X-ray afterglow~\citep{SWIFT,TrPi2017}, and had no observations hinting at a central engine remaining active following the GRB emission 
phase. Nevertheless, no electromagnetic observations rule out a longer-lived 
post-merger remnant for GW170817. 

Gravitational-wave emission mechanisms in this scenario include magnetic field-induced ellipticities~\citep{BoGo1996,Pa2001,Cut2002}, unstable bar modes~\citep{Lai1995,CoMe2009}, and unstable $r$-modes~\citep{LiOw1998,An1998}.  Estimates for the GW amplitude and detectability from such events vary across many orders of magnitude \citep[e.g.,][and Sec.~\ref{sec:Conclusion}]{CoMe2009,FaWu2013,DaGi2015,DoKo2015,LaGl2016}.

In summary, electromagnetic observations of this system do not provide definitive evidence for or against any of the four possible post-merger outcomes,
motivating this broad search using data-analysis algorithms which are robust to uncertain waveform morphologies.
We do not find any candidate post-merger GW signals associated with GW170817.
This is not surprising; even considering optimistic models of GW emission from the hypermassive or supramassive NS phases, the signal-to-noise ratio for a post-merger signal from $\sim$40\,Mpc in the current LIGO-Virgo network is less than $\sim$\,1-2 even for a matched-filter search \citep{2014PhRvL.113i1104T,ClBa2016}. However, we find that our current GW amplitude sensitivity is within approximately one order of magnitude of theoretical models for post-merger GW emission, implying that, with algorithmic improvements and the LIGO-Virgo network operating at design sensitivity~\citep{ObservingScenarios}, as well as future detectors, such emission might become detectable.

This paper is organized as follows.
In Sec.~\ref{sec:detchar} we describe the detectors and data set used.
In Sec.~\ref{sec:search} we present the search methods and results for both short- and intermediate-duration GW signals.
We discuss the implications and outlook for the future in Sec.~\ref{sec:Conclusion}.

\section{Detectors and data quality}
\label{sec:detchar}
\begin{figure*}
\includegraphics[width=\textwidth]{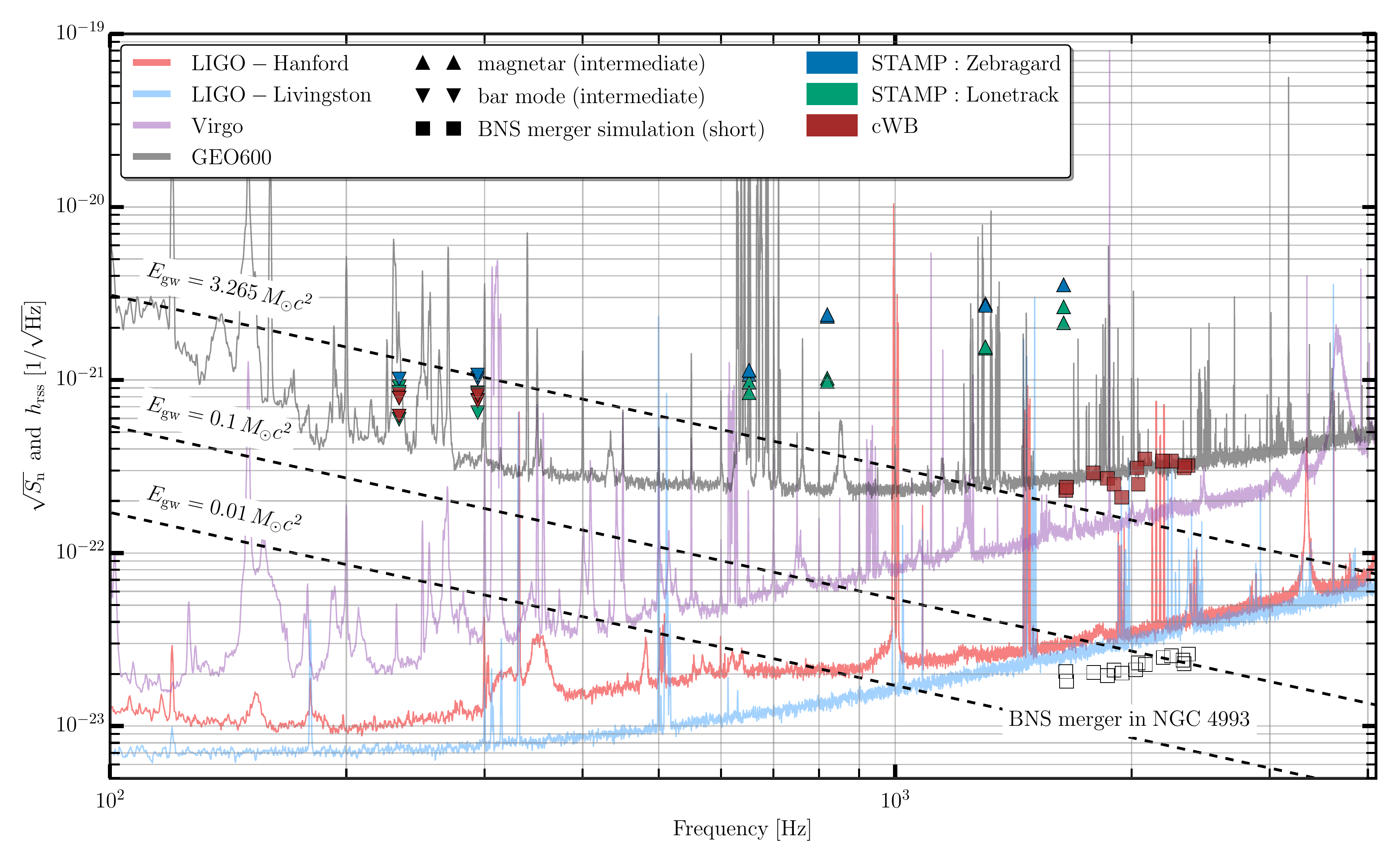}
\caption{
Noise amplitude spectral density $\sqrt{S_n}$ for the four GW detectors (solid curves), and detection efficiency root-sum-square strain amplitudes $h_{\rm rss}$ at 50\% false dismissal probability for various waveforms in the short- and intermediate-duration sensitivity studies.
The color code of the markers indicates the search, while the marker shapes correspond to the waveform families.
The red squares correspond to the short \code{cWB} analysis, the red triangles to the intermediate-duration \code{cWB} analysis, and the green and blue triangles correspond to the intermediate-duration \code{STAMP} Lonetrack and Zebragard analyses respectively.
The frequencies on the x-axis correspond to the average frequency of the injected waveform.
The short \code{cWB} analysis fixes the polarization to a value consistent with the pre-merger analysis \citep{DetectionPaper}, while for intermediate durations we marginalize over polarization (see text for details).
The top dashed black line indicates the maximum $h_{\rm rss}$ possible for a narrow-band \gw{} signal with fixed energy
content $E_{\mathrm{gw}}$, under the most optimistic assumption that the whole energy available after merger is radiated in \gw{s} at a certain frequency.
This energy is obtained from the pre-merger analysis~\citep{DetectionPaper} as $E_{\mathrm{gw}}=3.265\,\Msun c^2$ by subtracting the lower bound of $0.025\,\Msun c^2$ on the radiated energy from the upper end of the 90\% credible range on the total system mass.
The region above this line can thus be considered an unphysical part of parameter space, and in reality we expect \gw{} emission of only a fraction of this absolute upper bound -- as an example, lines at $0.1\,\Msun c^2$ and $0.01\,\Msun c^2$ are also shown.
The open squares represent the post-merger NR waveforms used in the short \code{cWB} analysis, but at the $h_{\rm rss}$ assuming the distance and orientation of GW170817 inferred from the pre-merger observation in~\citep{DetectionPaper}.  Figure produced using matplotlib~\citep{Hunter:2007}.
}
\label{fig:spectral_limits}
\end{figure*}

The LIGO~\citep{aLIGO}, Virgo~\citep{adVirgo}, and GEO600~\citep{GEO600} detectors were operating at the time of GW170817.  The noise amplitude spectral densities are shown in Fig.~\ref{fig:spectral_limits}, where the general trend of the detectors' sensitivities at high frequencies is due to the reduced interferometer response, interrupted by non-stationary spectral features, many of which have known origins.
The noise spectrum of LIGO Hanford is higher than that from Livingston in the frequency band from 100\,Hz to 1\,kHz;
one contribution is correlated laser noise that can be subtracted off-line (\citealt{P1700260}, used e.g. for the parameter estimation in \citealt{DetectionPaper}).
this search did not make use of such noise subtraction methods.
Virgo suffered from large noise fluctuations and non-stationary spectral features at frequencies above 2.5\,kHz~\citep{adVirgo}.

Due to a lack of detailed data quality studies available about GEO600, similar to those performed for LIGO and Virgo, data from that detector was not used in this analysis, although the sensitivity to a signal with time and sky location consistent with GW170817 would be roughly equal in Virgo and GEO600.
We note, however, that the network signal-to-noise ratio was dominated by the two LIGO detectors.

The algorithm used to search for short-duration signals (Coherent Wave Burst, \code{cWB}) used only LIGO data from 1024--4096\,Hz.
Two algorithms were used for intermediate-duration signals: The Stochastic Transient Analysis Multi-detector Pipeline (\code{STAMP}) searched from 24-2000\,Hz and 2000-4000\,Hz in LIGO-only data, while \code{cWB} searches from 24-2048\,Hz and used LIGO-Virgo data.
These algorithms are described in Sec.~\ref{sec:search}.

We whitened and removed stationary spectral lines of instrumental origin. 
Other techniques were employed to minimize the impact of non-stationary spectral features~\citep{O1LongDuration}. 
The data quality of the detectors was checked using the methods applied to previous gravitational-wave detections~\citep{Abbott:2016ctn}.
A short-duration instrumental disturbance occurred in the Livingston detector 1.1\,s before the coalescence time.  
Although this transient does not affect the performance of \code{cWB}, the \code{STAMP} analysis uses data in which the glitch is subtracted from the data \citep[see Fig.2 in][]{DetectionPaper}.

LIGO's calibration uncertainty is 7\% in amplitude and 3\,degrees in phase below 2\,kHz  \citep{DetectionPaper}, and 8\% in amplitude and 4\,degrees in phase above 2\,kHz~\citep{Cahillane:2017jb}.
Virgo's calibration uncertainty is 10\% in amplitude and 10\,degrees in phase up to 5\,kHz~\citep{DetectionPaper}.  Calibration uncertainties are not taken into account in calculations of upper limits.

\section{Search Methods and Detection Efficiencies}
\label{sec:search}
In situations with great theoretical uncertainties, where no complete set of accurate GW template waveforms is available, a matched-filter search is not feasible.
Instead, an efficient solution is to search for excess power in spectrograms (also called frequency-time or $ft$-maps) of GW detector data \citep{AnBr2000,KlMi2004}. Pattern recognition algorithms are used to identify the presence of GW signals in these maps \citep{ThCo2013,SuJo2010,ThKa2011,CoLi2015,KlVe2016}.
Here, to account for the large uncertainty in the nature of the remnant, we employ a number of algorithms, each designed to coherently combine data from multiple GW detectors, with different data-processing and clustering techniques that make them respond differently to different waveform models.
These algorithms are designed to be sensitive to a wide variety of signal morphologies, and while we test their sensitivity to a number of post-merger waveform models, they are designed so as to be robust against the significant theoretical uncertainties by using generic clustering schemes.
Each algorithm performs the search at a single sky position, which we take to be the direction of the host galaxy for the optical counterpart, NGC4993~\citep[RA = 13.1634\,hrs, Dec. = $-23.3815^\circ$;][]{CoFo2017,MMAPaper}.
Below, we briefly describe each algorithm used in this search and present their findings.

\subsection{Short duration ($\lesssim1$\,s) signals}\label{sec:short}
We perform an analysis targeting short-duration, high-frequency \gw{s} near the time of coalescence designed to be sensitive to unmodeled signals. 
This search for \gw{} bursts is performed using the \code{cWB} algorithm~\citep{KlVe2016}.
We search for statistically significant coherent excess power due to \gw{} bursts
in a  2\,s long window which begins at 1187008882 GPS time, includes the estimated time of coalescence, and extends forward in time covering the entire delay between the merger and the GRB~\cite[\FermiDelay \,s,][]{GRBGWPaper}.

The \code{cWB} algorithm performs a maximum likelihood evaluation of coherent excess power in a multi-resolution Wilson-Daubechies-Meyer wavelet transform, which is performed on the strain from each detector~\citep{KlVe2016}. The analysis ranks candidate events by their coherent network signal-to-noise ratio.  Statistical significance of candidate events is found by comparing the ranking statistic with a background distribution measured from 5.6 days of coincident data from Livingston and Hanford during the period 13--21 August. This data is ``time-shifted,'' which means that a non-physical time lag is introduced between the detector analyzed so as to remove correlated gravitational-wave signals. This data is also ``off-source,'' which means it is outside of the 2\,s window over which gravitational waves are searched for. This analysis yields an estimate of the false-alarm probability for a given possible detection.

We search over a frequency range of 1024--4096\,Hz.
No significant events are found within
the 2\,s ``on-source'' window.
The sensitivity of the \code{cWB} analysis is characterized through Monte-Carlo simulations in which waveforms from binary NS post-merger simulations are added to data from off-source periods (see Appendix~\ref{app:bnssims} for details). The simulated sources are placed at the known sky-location of the optical counterpart of GW170817 and with orbital inclination consistent with the pre-merger analysis~\citep{DetectionPaper}.  
The waveform amplitudes are varied to determine the efficiency as a function of signal strength; see~\citet{O1LongDuration} for an expanded discussion.  
The response of a given detector to the impinging \gw{} is assumed to take the form

\begin{equation}
\label{eq:response}
s(t) =  F_+(\Theta,\psi) h_+(t) + F_{\times}(\Theta,\psi) h_{\times}(t), 
\end{equation}
where $F_+$ and $F_{\times}$ are the antenna patterns for a given detector, $\Theta$ encodes the direction to the source and $\psi$ is the polarization angle.

It is customary to express the sensitivity of a search to a given model waveform in $h_{\rm rss}^{50\%}$, which is the root-sum-squared strain amplitude of signals which are detected with 50\% efficiency \cite{O1LongDuration}.
The detection criterion has been chosen in this specific search by setting a detection threshold on the significance of candidates which corresponds to a false-alarm probability of $10^{-4}$.  The quantity $h_{\mathrm{rss}}$ is defined as
\begin{equation}\label{eq:hrss}
h_{\mathrm{rss}} = \sqrt{2\int_{f_{\textrm{min}}}^{f_{\textrm{max}}} \left( |\tilde{h}_+(f)|^2 + |\tilde{h}_{\times}(f)|^2 \right) \diff f},
\end{equation}
where $f_{\mathrm{min}}$ and $f_{\mathrm{max}}$ are respectively the minimum and maximum frequencies over which the search is performed.
The search sensitivities are shown in Fig.~\ref{fig:spectral_limits} and Table~\ref{tab:bns} in terms of the average frequency of each waveform, $\bar{f}$. 
We also provide as a point of comparison the $h_{\rm rss}$ of the same NR waveforms used in the analysis but assuming the distance of GW170817.  It is worth noting that softer \eos{s} which lead to more compact stars exhibit a longer duration, higher frequency inspiral phase~\citep{2013ApJ...773...78B} and a more dense remnant with relatively high frequency post-merger oscillations.  All NR waveforms have dominant emission well-above 1\,kHz, however searching from $f_{\mathrm{min}}=1024$\,Hz is a conservative choice made to avoid missing any post-merger signal content from stiff \eos{s} but permits pre-merger and merger-signal content from soft \eos{s}.  In the end, the search finds no evidence for any \gw{} signal in this band and the waveforms used to form upper limits are dominated by the postmerger phase, although they do allow for some part of the late inspiral and merger.

The strains required to produce a 50\% probability of signal detection lie between $2.1\times 10^{-22}$\,Hz$^{-1/2}$ and $3.5\times 10^{-22}$\,Hz$^{-1/2}$.  The \gw{} energy radiated by an isotropically emitting source is given by \cite{2013arXiv1304.0210S}

\begin{align}\label{eq:energy}
E_{\mathrm{gw}}^{\mathrm{iso}} & =  \frac{\pi c^3}{2 G} \dist^2 \int   \mathrm{d}\Omega \int_{f_{\textrm{min}}}^{f_{\textrm{max}}} \mathrm{d}f~f^2 \left(|\tilde{h}_+(f)|^2 + |\tilde{h}_{\times}(f)|^2 \right) \nn \\
 & \approx \frac{\pi^2 c^3}{G} \dist^2 \bar{f}^2 h_{\mathrm{rss}}^2,
\end{align}
where $\dist$ is the distance to the source.
Using the $h_{\rm rss}^{50\%}$ sensitivities described above, we find that the energies to which the search is sensitive are 4.8--19.6\,$\Msun c^2$, where the range corresponds to the variety of waveforms used. We are therefore not able to constrain post-merger emission from a possible hypermassive NS associated with GW170817.

A separate analysis of the LIGO-Virgo data for unmodelled short duration bursts within a [-600,+60] second window around GRB170817A is reported in \cite{GRBGWPaper} using the X-Pipeline package \citep{SuJo2010,WaSu2012}. This analysis searched the frequency band 20-1000\,Hz. The inspiral phase of GW170817 was detected with a significance of 4.2\,$\sigma$, rising to 5\,$\sigma$ when the analysis is constrained to the optical counterpart location. However, no significant events were found following the merger. Limits on the amplitude of \gw{} emission below 1000\,Hz are consistent with those reported here.

\subsection{Intermediate duration ($\lesssim500$ s) signals}\label{sec:intermediate}

For intermediate-duration signals, we employ search algorithms adapted from the all-sky searches described in \cite{O1LongDuration}. The main difference with respect to the all-sky searches is that instead of searching over many possible sky positions, we again use the known sky position of the optical counterpart. Together with the limited time range to search over, this effectively reduces the number of accidental coincident triggers. Two algorithms are employed: \code{STAMP} and \code{cWB}.

While the algorithms are sensitive to rather general waveform morphologies, we test the efficiency of signal recovery for both by a set of specific waveform models to determine $h_{\rm rss}^{50\%}$.
We coherently add these simulated post-merger signals to the data of LIGO Hanford and Livingston covering the on-source period.
The waveforms' polarizations are allowed to vary uniformly in $\psi$ and $\cos\iota$, which corresponds to selecting from an isotropic distribution.
The sky positions are fixed to the position of the optical transient.
We describe the
waveform models in the next section.

\subsubsection{Waveform Models}\label{sec:WFs}
Two types of physically-motivated waveform morphologies are considered, corresponding to GWs either from secular bar modes~\citep{Lai1995} or caused by magnetic-field induced ellipticities of the nascent star~\citep[referred to as \emph{magnetar} waveforms in the following]{Cut2002}.
Another interesting emission mechanism are unstable $r$-modes~\citep{An1998,MyCo2015}; we do not use such waveforms here due to the duration of their emission, and so searches covering significantly longer timescales will be required.

The secular bar mode is a GW-driven instability~\citep{Cha1970,FrSc1975}, where the growth timescale of the mode is determined by the ratio of kinetic to binding energy of the star~\citep{Lai1995}.  The corresponding waveforms~\citep{CoMe2009} and specific parameters of the model used for the waveforms are given in Appendix~\ref{app:barmode} and Table~\ref{tab:barmode}.

The magnetar waveforms assume that the merger results in a star that is rapidly spinning down, whose internal magnetic field has been wound up, generating significant stellar ellipticity~\citep[e.g.,][]{Cut2002,DaSh2009,CiRe2013}.  The star then undergoes a spin-flip instability, causing it to become an orthogonal rotator, and hence maximal emitter of GWs.  The specific waveform model is derived in~\citet{LaSa2017}.  
The waveform is parameterized by four parameters: a braking index, stellar ellipticity, the initial GW frequency, and the spindown timescale.
Details of these waveforms and their parameters are given in Appendix~\ref{app:magnetar} and Table~\ref{tab:magnetar}.

\subsubsection{STAMP}\label{sec:longSTAMP}
\code{STAMP} employs spectrograms with 1\,s $\times$ 1\,Hz pixels created from the cross-correlation of data between spatially separated detectors \citep{ThKa2011}, which in this case are LIGO Hanford and LIGO Livingston.  We use
500\,s spectrograms covering two frequency bands: 24-2000\,Hz and 2000-4000\,Hz.
The on-source data from the time of merger to the end of the second observing run is split into these 500\,s spectrograms with 250\,s overlap between them.
The time-shifted off-source data is taken from August 3, 2017 until the time of the merger.
These are searched with both a seed-based clustering method (Zebragard) and with a seedless pattern-recognition algorithm (Lonetrack) that integrates the pixels across tracks which are picked randomly from a large set of B\'ezier templates over the spectrogram \citep{ThCo2013,ThCo2015}.

\subsubsection{Coherent Wave Burst}\label{sec:longcWB}
For the intermediate-duration search, the \code{cWB} algorithm (see Sec.~\ref{sec:short} for algorithm details) searches between 24-2048\,Hz using LIGO-Virgo data from the time of the merger to 1000\,s later.  Selection criteria for candidate gravitational wave triggers are based on the duration of the signal reconstructed by the algorithm as described in \citet{O1LongDuration}.
The on-source time window is taken to be the time of the merger until 1000\,s later, while the off-source data is the period from 13--21 August with the data time-shifted such that no coherent signals remain.

\subsubsection{Results}\label{sec:intresults}

In the \code{STAMP} analysis, the triggers found in the on-source period are compared to the estimated background of accidental coincident triggers; there is no significant excess of coherent events during this time period in the 24-4000\,Hz frequency range searched corresponding to a $10^{-2}$ false-alarm probability.
Similarly, no GW transient candidates have been found by \code{cWB} above a ranking statistic value corresponding to a $10^{-4}$ false-alarm probability in the frequency band 24-2048\,Hz.

We determine detection efficiencies for the models considered in Sec.~\ref{sec:WFs}. For \code{STAMP}, we report the equivalent energy released at which the algorithms recover 50\% or more of the injected signals at a false-alarm probability of $10^{-2}$, as well as the corresponding $h_{\rm rss}$. For \code{cWB}, we report the results at a false-alarm probability of $10^{-4}$. Due to the rapid rise in background events between false alarm probabilities of $10^{-2}$ and $10^{-4}$, the results presented here do not depend strongly on this choice. These false alarm probabilities were chosen as they correspond to a false alarm rate of approximately 1 per year. 
The best \code{STAMP} results correspond to an $h_{\rm rss}^{50\%}=\barmodesFLonetrack{}\times10^{-22}$\,Hz$^{-1/2}$ with equivalent energy of $E_\mathrm{GW}=\barmodesULSTAMP{}$ for the bar-mode signal models.
The results for the \code{cWB} analysis are similar.
For the magnetar signal models, the best \code{STAMP} results are $h_{\rm rss}^{50\%}=\msmagnetarBLonetrack{}\times10^{-22}$\,Hz$^{-1/2}$ and $E_\mathrm{GW}=\magnetarULSTAMP{}$.
\code{cWB} did not analyze these waveforms. The energy limits are computed using Eq.~\ref{eq:energy} where isotropic GW emission is assumed and the $h_{\rm rss}^{50\%}$ values are marginalized over polarization.

\section{Implications and Conclusions}
\label{sec:Conclusion}

We report on a search for GWs from the post-merger remnant following the binary NS coalescence GW170817,
using robust and generic time-frequency excess power analysis methods.
Such GWs can come from a short-lived hypermassive NS lasting $\lesssim1$ s before collapsing to a BH or from a longer-lived supramassive or stable NS.
We find no evidence in our data for GWs after the merger of GW170817.
If a signal exists, it is too weak to be detected with current sensitivity and analysis algorithms.

For the data set and methods employed in this paper, we find search sensitivities, in terms of GW signal amplitude, that are approximately an order of magnitude from expectations for GW emission in the literature.  For example, short-lived hypermassive NSs are expected to emit a few percent of a solar mass in gravitational-wave energy \citep[e.g., see the two lower dashed lines in Fig.~\ref{fig:spectral_limits} that represent 1\% and 10\% of a solar mass;][]{KiSe2009,ClBa2014,BeNa2015,EnCi2016,DiBe2017,DiUj2017b,Feo2017}, while our minimum 50\% efficiency is $E_{\rm GW}\lesssim\,4.8\,\Msun c^2$ (see Sec.~\ref{sec:short}).  
Gravitational-wave emission from a representative sample of these numerical relativity simulations are shown as open squares in Fig.~\ref{fig:spectral_limits}, which are approximately an order of magnitude in strain below the $h_{\rm rss}^{50\%}$ points for the corresponding waveforms (filled squares).
For intermediate-duration signals from supramassive or stable NSs (Sec. \ref{sec:intermediate}), we find a minimum of $E_{\rm GW}\lesssim$\,\magnetarULSTAMP{} for the millisecond magnetar model~\citep{LaSa2017}, and $E_{\rm GW}\lesssim$\,\barmodesULSTAMP{} for the model describing secular bar modes~\citep{CoMe2009}.  

Figure~\ref{fig:spectral_limits} shows the $h_{\rm rss}^{50\%}$ for the considered waveform models as a function of the waveform's signal-weighted frequency.  Based on this, the distance of 40\,Mpc for the binary NS is approximately an order of magnitude greater than the distances to which we are sensitive. This gives significant motivation to continue searching for intermediate-duration post-merger remnants in later iterations of the advanced (or future) GW detectors as well as the development of improved analysis methods.

GW170817 was detected in the second observing run of the advanced GW detectors. Further improvements towards their design sensitivity are now underway~\citep{ObservingScenarios}.
At design sensitivity, a matched-filter search with precisely modelled post-merger waveforms could detect signals from a hypermassive NS remnant out to distances of $\sim 20\mbox{--}40$\,Mpc (computed as the single-detector horizon distance for a signal-to-noise threshold of 5) \citep{2014PhRvL.113i1104T,ClBa2016}.
Conversely, with current detector sensitivities and the more robust search methods not relying on matched filtering that are employed in this paper, a post-merger detection for GW170817 is not very likely a priori, but the theoretical uncertainties still make a search important.
By using algorithms designed to be sensitive to generic signals, the searches are robust to these theoretical uncertainties and capable of detection of unmodeled signals.

This study motivates increased research and development towards improved sensitivity at high-frequencies in current instruments, planned upgrades and also third-generation interferometers.  Future improvements can also be made to the search methods presented in this paper.
For example, in searching for short-duration signals, the sparsity of numerical-relativity waveforms makes it challenging to perform matched-filter searches for the inspiral, merger and hypermassive phases.
However, the dominant post-merger GW modes are dictated by only a few pre-merger parameters and the equation of state~\citep[e.g.,][]{ReBa2013,BeDi2015,BaSt2015,ReTa2016}, implying more sensitive techniques can be developed that use parameters measured during the inspiral to inform priors on the physical parameters of the post-merger remnant.

A full matched-filter search is likely not computationally possible for intermediate-duration signals due to the large parameter spaces and theoretical uncertainties involved with the waveforms.
However, sensitivity can be improved by targetting specific emission models \citep[e.g.,][]{CoCo2016}.

A search for longer-duration ($\gtrsim$\,1\,day) remnant signals is also planned, with the maximum detectable signal length limited by the 7.9 days of data available following the merger until the official end of the LIGO-Virgo second observing run.
A variant of the intermediate-duration algorithms with different pixel sizes can be used to create spectrograms that cover the full duration of the analysis, making it more sensitive to longer-lived signals than the maps employed here \citep{ThMa2015}.
Moreover, a variety of methods have been developed to search for persistent, nearly-monochromatic signals from mature NSs~\citep[see][for reviews]{Prix:2009oha,Riles:2012yw,Bejger2017}.
Several of these could be modified to search for a long-lived post-merger signal,
though the expected rapid decrease in frequency is likely to pose technical challenges to current algorithms.

In addition to improving the sensitivity to potential post-merger signals of GW170817, another important program is to improve our ability to detect post-merger GWs from future LIGO/Virgo discoveries of binary NS mergers.  At design sensitivity, Advanced LIGO and Advanced Virgo both aim to be approximately a factor three better in broadband sensitivity than during the second observing run~\citep{ObservingScenarios}, and next generation detectors will improve the sensitivities significantly beyond that.  This provides a number of opportunities.  Clearly, increased sensitivity in the $\gtrsim$ kHz range implies improved ability to detect single post-merger signals.  Moreover, increased broadband sensitivity implies higher rates of binary NS inspiral and merger detections, and hence might make possible power or coherent stacking of events to increase our sensitivity to post-merger physics~\citep{BoCh2017,YaPa2017}.

\acknowledgements
The authors gratefully acknowledge the support of the United States
National Science Foundation (NSF) for the construction and operation of the
LIGO Laboratory and Advanced LIGO as well as the Science and Technology Facilities Council (STFC) of the
United Kingdom, the Max-Planck-Society (MPS), and the State of
Niedersachsen/Germany for support of the construction of Advanced LIGO 
and construction and operation of the GEO600 detector. 
Additional support for Advanced LIGO was provided by the Australian Research Council.
The authors gratefully acknowledge the Italian Istituto Nazionale di Fisica Nucleare (INFN),  
the French Centre National de la Recherche Scientifique (CNRS) and
the Foundation for Fundamental Research on Matter supported by the Netherlands Organisation for Scientific Research, 
for the construction and operation of the Virgo detector
and the creation and support  of the EGO consortium. 
The authors also gratefully acknowledge research support from these agencies as well as by 
the Council of Scientific and Industrial Research of India, 
the Department of Science and Technology, India,
the Science \& Engineering Research Board (SERB), India,
the Ministry of Human Resource Development, India,
the Spanish  Agencia Estatal de Investigaci\'on,
the  Vicepresid\`encia i Conselleria d'Innovaci\'o, Recerca i Turisme and the Conselleria d'Educaci\'o i Universitat del Govern de les Illes Balears,
the Conselleria d'Educaci\'o, Investigaci\'o, Cultura i Esport de la Generalitat Valenciana,
the National Science Centre of Poland,
the Swiss National Science Foundation (SNSF),
the Russian Foundation for Basic Research, 
the Russian Science Foundation,
the European Commission,
the European Regional Development Funds (ERDF),
the Royal Society, 
the Scottish Funding Council, 
the Scottish Universities Physics Alliance, 
the Hungarian Scientific Research Fund (OTKA),
the Lyon Institute of Origins (LIO),
the National Research, Development and Innovation Office Hungary (NKFI), 
the National Research Foundation of Korea,
Industry Canada and the Province of Ontario through the Ministry of Economic Development and Innovation, 
the Natural Science and Engineering Research Council Canada,
the Canadian Institute for Advanced Research,
the Brazilian Ministry of Science, Technology, Innovations, and Communications,
the International Center for Theoretical Physics South American Institute for Fundamental Research (ICTP-SAIFR), 
the Research Grants Council of Hong Kong,
the National Natural Science Foundation of China (NSFC),
the Leverhulme Trust, 
the Research Corporation, 
the Ministry of Science and Technology (MOST), Taiwan
and
the Kavli Foundation.
The authors gratefully acknowledge the support of the NSF, STFC, MPS, INFN, CNRS and the
State of Niedersachsen/Germany for provision of computational resources.

\appendix
\section{Waveforms}
\label{sec:WFapp}
Here we provide details of the waveform models used to determine the detection efficiency of our search algorithms.  For short-duration signals (Sec.~\ref{sec:short}) we use simulations of binary NS mergers.  For intermediate durations (Sec.~\ref{sec:WFs}), we use two models: secular bar modes and magnetar waveforms.

\subsection{Binary neutron star waveforms}\label{app:bnssims}
We determine the efficacy of the short-duration \code{cWB} analysis (Sec.~\ref{sec:short}) through Monte-Carlo simulations using \gw{} waveforms derived from simulations of binary NS systems that include a post-merger phase.  Table~\ref{tab:bns} provides a summary of the waveforms in terms of their expected $h_{\mathrm{rss}}$ for a binary NS at 40 Mpc and the $h_{\mathrm{rss}}$ required for 50\% detection efficiency by \code{cWB} with a $10^{-4}$ false-alarm probability, $h_{\mathrm{rss}}^{50\%}$.   
 
\begin{table}[h!]
\centering
\label{tab:bns}
\begin{tabular}{lclcccc}
\hline \hline
Equation of state & $m_1$ & $m_2$ & $\bar{f}$ & Simulation  & $h_{\rm rss}$ (expected) & $h_{\rm rss}^{50\%}$  \\
    & [M$_{\odot}$] & [M$_{\odot}$] &  [Hz] &  & [$10^{-22}/\sqrt{\rm Hz}$]  & [$10^{-22}/\sqrt{\rm Hz}$]  \\
 \hline
 H4~\citep{PhysRevLett.67.2414}    & 1.25 & 1.25 & 1946 & \citet{2015PhRvD..91f4001T} & $0.21$ & $2.1$ \\
 H4   \citep{PhysRevLett.67.2414}   & 1.3  & 1.3 & 2083 & \citet{2015PhRvD..91f4001T} & $0.23$ & $3.5$ \\
 H4   \citep{PhysRevLett.67.2414}  & 1.35 & 1.35 & 2247 & \citet{CiKa2017}  & 0.26 & 3.4  \\
 H4   \citep{PhysRevLett.67.2414}  & 1.42 & 1.29 & 2192 & \citet{CiKa2017}  & 0.26 & 3.4 \\
 H4   \citep{PhysRevLett.67.2414}  & 1.54 & 1.26 & 2030 & \citet{KaGi2016} & 0.22 & 3.1\\
 LS220~\citep{1991NuPhA.535..331L} & 1.20 & 1.50 & 1900 & \citet{2013PhRvL.111m1101B}  & $0.22$ & $2.5$   \\
 SHT~\citep{PhysRevC.82.015806}    & 1.40 & 1.40 & 1788 & \citet{PhysRevD.96.043019}  & $0.21$ & $2.9$  \\
 SFHx~\citep{2013ApJ...774...17S}  & 1.2 & 1.5 & 1650 & \citet{2013PhRvL.111m1101B}   & $0.21$ & $2.3$  \\
 SFHx~\citep{2013ApJ...774...17S}  & 1.35 & 1.35 & 2040 & \citet{2013PhRvL.111m1101B}  & $0.24$ & $2.5$  \\
 SLy~\citep{SLyEoS}   & 1.25 & 1.25 & 2333 & \citet{2015PhRvD..91f4001T} & $0.23$ & $3.2$  \\
 SLy~\citep{SLyEoS}   & 1.3 & 1.3 & 2325 &  \citet{2015PhRvD..91f4001T} & $0.25$ & $3.1$  \\
 SLy~\citep{SLyEoS}   & 1.35 & 1.25 & 2363 & \citet{2015PhRvD..91f4001T} & $0.27$ & $3.2$  \\
 TMA~\citep{1995NuPhA.588..357T}   & 1.20 & 1.50 & 1864 & \citet{2013PhRvL.111m1101B}  & $0.19$ & $3.2$  \\
 TMA~\citep{1995NuPhA.588..357T}   & 1.35 & 1.35 & 1653 & \citet{2013PhRvL.111m1101B}  & $0.20$ & $2.4$ \\ 
  \hline
\end{tabular}
\caption{Sensitivity of the \code{cWB} pipeline to waveforms generated by binary NS simulations.  Waveforms were selected to represent a variety of equations of state (first column) and progenitor mass configurations (second and third columns).  The fourth column is the mean frequency for each waveform $\bar{f}$, and the fifth column is the reference for the BNS simulation.  The sixth column is the root-sum-squared strain $h_{\rm rss}$ predicted by that simulation for a post-merger signal from a BNS with distance and inclination consistent with estimates from the inspiral analysis~\citep{DetectionPaper}.  The seventh column shows the $h_{\rm rss}$ required for 50\% detection efficiency with a false-alarm probability of $10^{-4}$, $h_{\rm rss}^{50\%}$. }
\end{table}

\subsection{Secular bar mode waveforms}\label{app:barmode}
Long-lived post-merger remnants may be unstable due to the secular bar-mode instability.  This instability occurs when the ratio of rotational kinetic energy to gravitational binding energy $T/|W|$ is in the range $0.14 < T/|W|< 0.27 $~\citep{Lai1995}.  For all injected waveforms used in this study we follow the treatment described in \citet{CoMe2009}, which we briefly summarize. We set $T/|W|=0.2$ for the kinetic-to-gravitational potential energy ratio of the initial axisymmetric configuration (in the middle of the secular instability range).  The NS spin-down is then determined by the combination of magnetic dipole and GW losses \citep{CoMe2009}:
\begin{align}
	\frac{dE}{dt}=-\frac{B^2R^6\Omega_{\rm eff}^4}{6c^3}-\frac{32GI^2\epsilon^2\Omega^6}{5c^5}.\label{eq:spindown}
\end{align}
Here, $B$ is the star's dipolar field strength at the poles, $R$ is the mean stellar radius (i.e. the geometric mean of the principal axes of the star), $\Omega$ is the star's angular frequency, $\Omega_{\rm eff}$ is the effective angular frequency (which includes the effect of internal fluid motions), $\epsilon$ is the ellipticity, and $I$ is the moment of inertia with respect to the rotation axis.  The GW strain is then
\begin{align}
	h(t)=\frac{4G}{c^4}\frac{I\epsilon \Omega^2}{\dist},\label{eq:ht}
\end{align}
where $\Omega$ is found by integrating Eqn.~(\ref{eq:spindown}).

For all injected waveforms we assume a total NS mass of $2.6\,\Msun$, which is close to the lower bound of the estimated total mass range for GW170817 \citep[$2.73\,$M$_{\odot}$; see][]{DetectionPaper} and to the lower bound of $2.57$\,M$_{\odot}$ for the total mass of other known binary NS systems \citep[see][and references therein]{DetectionPaper}.  For a given initial $T/|W|$ value, the total radiated energy during the secular bar-mode evolution scales as $M^2$ \citep[see e.g. Fig. 3 in][]{Lai1995}, implying that our value of the mass can be regarded as conservative within the (optimistic) assumption that all of the total mass of the binary goes into the post-merger remnant.
We use a range of magnetic field values from $10^{13}$ to $5\times10^{14}$\,G.  Fields higher than $\sim 10^{15}$\,G result in rapid spindown of the star, and hence uninteresting GW amplitudes; fields lower than $\sim10^{13}$\,G are unrealistic for such systems given the post-merger remnant dynamics that wind up strong fields.  Because we do not know the ultimate fate of the bar-shaped remnant, and whether it can survive up to the ultimate Dedekind configuration \citep{Lai1995}, we only evolve waveforms up to a time when the luminosity emitted in GWs is 1\% of the peak value, which is sufficient to capture the bulk of the energy emitted in GWs. Table~\ref{tab:barmode} shows the specific parameters used for our waveforms, as well as $h_{\rm rss}$ at 50\% efficiency computed at a fixed false alarm probability for each of the pipelines used in this search.  These results are also shown in Fig.~\ref{fig:spectral_limits}.  

\begin{table}[h!]
\centering
\label{tab:barmode}
\begin{tabular}{ccccc|ccc}
	\hline\hline
	\multicolumn{5}{c}{Properties} & \multicolumn{3}{c}{$h_{\rm rss}^{50\%}$ [$10^{-22}/\sqrt{\rm Hz}$]}\\
	\hline
	$R$ & $B$ & $T$ & $f_0$ & $f_f$ & \code{cWB} & \multicolumn{2}{c}{\code{STAMP}}\\
	{ [km]} & [G] & [s] & [Hz] & [Hz] & & Lonetrack & Zebragard\\
	\hline\hline
	12 & $10^{13}$ & 277 & 449 & 139 & \barmodesAcWB & \barmodesALonetrack & \barmodesAZebragard\\
	12 & $10^{14}$ & 237 & 449 & 139 & \barmodesBcWB & \barmodesBLonetrack & \barmodesBZebragard\\
	12 & $5\times10^{14}$ & 107 & 449 & 139 & \barmodesCcWB & \barmodesCLonetrack & \barmodesCZebragard\\
	14 & $10^{13}$ & 509 & 356 & 111 & \barmodesDcWB & \barmodesDLonetrack & \barmodesDZebragard\\
	14 & $10^{14}$ & 396 & 356 & 111 & \barmodesEcWB & \barmodesELonetrack & \barmodesEZebragard\\
	14 & $5\times10^{14}$ & 136 & 356 & 111 & \barmodesFcWB & \barmodesFLonetrack & \barmodesFZebragard\\
	\hline
\end{tabular}
\caption{$h_{\rm rss}$ at 50\% efficiency to the bar mode waveforms computed for a false-alarm probability of $10^{-2}$ for \code{STAMP} and $10^{-4}$ for \code{cWB}---see Sec.~\ref{sec:intermediate}. Here, $B$ is the star's dipolar magnetic field strength at the pole, $R$ the mean stellar radius, $T$ the duration of the waveform in seconds, and $f_0$ and $f_f$ define the beginning and end of the frequency range where the bulk of the GW energy is emitted. Please see Sec.~\ref{app:barmode} for further waveform details.}
\end{table}

\subsection{Magnetar waveforms}\label{app:magnetar}
Gravitational waveforms from spinning-down nascent NSs with arbitrary braking index are derived in \citet{LaSa2017}.  Here we assume that the rotational evolution of the star is described by the torque equation: $\dot{\Omega}\propto\Omega^n$, where $n$ is the braking index.  Integrating the torque equation enables one to derive the star's spin evolution, and hence the GW frequency
\begin{align}
	f(t)=f_0\left(1+\frac{t}{\tau}\right)^{1/(1-n)},
\end{align}
where $f_0$ is the initial GW frequency, and $\tau$ is the spindown timescale.  The gravitational-wave strain is then given by Eqn.~(\ref{eq:ht}), where the GW frequency is twice the star's spin frequency.

A braking index of $n=5$ represents gravitational-wave driven spindown due to stellar ellipticity $\epsilon$, whereas an unchanging dipolar magnetic field in vacuum induces a braking index of $n=3$.  Observations of X-ray afterglows from short GRBs allow for constraints on $\tau$ and $\Omega(t=0)$, and hence $f_0$~\citep{Row2013}, as well as $\epsilon$~\citep{LaGl2016}, and $n$~\citep{LeLe2017}.  The braking index of two milllisecond magnetars have been measured, both below $3$; we therefore choose $n=2.5$ and $n=5$ to adequately sample the space.  Empirically, such stellar ellipticities are limited to $\lesssim10^{-2}$, although theoretically such a large value is difficult to generate with internal magnetic fields as it requires a field of $\sim10^{17}$ G.  Large-scale $\alpha$--$\Omega$ dynamos may generate internal fields of $\sim10^{16}$\,G~\citep{ThDu1993}, with small-scale turbulent dynamos potentially amplifying the field to $\lesssim5\times10^{16}$\,G~\citep{ZrMa2013}, implying ellipticities as high as $\approx2.5\times10^{-3}$~\citep{Cut2002}.  In principle, the ellipticity should effect the GW frequency evolution, however in these waveform models that factor is absorbed into $\tau$. Finally, while X-ray observations indicate $10\lesssim\tau/{\rm s}\lesssim10^5$~\citep{Row2013}, we only show waveform results here using $\tau=100$ s as larger $\tau$ yields uninteresting GW limits.  

Table~\ref{tab:magnetar} shows waveform parameters used in this study and corresponding $h_{\rm rss}$ at 50\% efficiency computed for a fixed false alarm probability of $10^{-2}$.  We do not list the duration of the waveform as all magnetar waveforms are longer than the search duration.

\begin{table}[h!]\label{tab:magnetar}
\centering
\begin{tabular}{cccc|cc}
	\hline\hline
	\multicolumn{4}{c}{Properties} & \multicolumn{2}{c}{$h_{\rm rss}^{50\%}$ [$10^{-22}/\sqrt{\rm Hz}$]}\\
	\hline
	$\epsilon$ & $n$ & $f_0$  & $f_{500}$ & \multicolumn{2}{c}{\code{STAMP}}\\
	{} & & [Hz] & [Hz] & Lonetrack & Zebragard\\
	\hline\hline
	0.01 & 2.5 & 1000 & 303 & \msmagnetarALonetrack & \msmagnetarAZebragard \\
	0.001 & 2.5 & 1000 & 303 & \msmagnetarBLonetrack & \msmagnetarBZebragard \\
	0.01 & 5 & 1000 & 639 & \msmagnetarCLonetrack & \msmagnetarCZebragard \\
	0.001 & 5 & 1000 & 639 & \msmagnetarDLonetrack & \msmagnetarDZebragard \\
	0.01 & 2.5 & 2000 & 606 & \msmagnetarILonetrack & \msmagnetarIZebragard \\
	0.001 & 2.5 & 2000 & 606 & \msmagnetarJLonetrack & \msmagnetarJZebragard \\
	0.01 & 5 & 2000 & 1278 & \msmagnetarKLonetrack & \msmagnetarKZebragard \\
	0.001 & 5 & 2000 & 1278 & \msmagnetarLLonetrack & \msmagnetarLZebragard \\
	\hline
\end{tabular}
\caption{$h_{\rm rss}$ at 50\% efficiency to the magnetar waveforms computed for a false-alarm probability of $10^{-2}$ for the \code{STAMP} pipelines used in the intermediate-duration search---see Sec.~\ref{sec:intermediate}. The first four columns are respectively the stellar ellipticity $\epsilon$, the braking index $n$, the initial  \gw{} frequency $f_0$, and the \gw{} frequency after 500 s $f_{500}$, which is the duration of the analyses. Please see Sec.~\ref{app:magnetar} for further waveform details.}
\end{table}

\bibliographystyle{aasjournal}
\bibliography{references}

\iftoggle{endauthorlist}{
  %
  %
  \let\author\myauthor
  \let\affiliation\myaffiliation
  \let\maketitle\mymaketitle
  \title{Authors}
  
  \newpage
  \maketitle
}

\end{document}